\documentclass[twocolumn,showpacs,preprintnumbers,amsmath,amssymb, prb]{revtex4}
\usepackage{graphicx}
\usepackage{dcolumn}
\usepackage{bm}
\begin{document}

\preprint{APS/123-QED}
\title{Magnetic properties of the heavy fermion antiferromagnet CeMg$_3$}
\author{Pranab Kumar Das, Neeraj Kumar,  R. Kulkarni and A. Thamizhavel}
\affiliation{Department of Condensed Matter Physics and Materials
Science, Tata Institute of Fundamental Research, Homi Bhabha Road,
Colaba, Mumbai 400 005, India.}
\date{\today}

\begin{abstract}
We have grown the single crystals of  CeMg$_3$ and its non-magnetic analogue LaMg$_3$,  which crystallize in the cubic crystal structure with the space group \textit{Fm\rm{\={3}}m}, and studied their magnetic properties  on well oriented single crystals by measuring the magnetic susceptibility, magnetization, electrical resistivity and heat capacity.  CeMg$_3$ orders antiferromagnetically with a N\'{e}el temperature $T_{\rm N}$ of  2.6~K.  The specific heat capacity  at low temperature exhibits an enhanced Sommerfeld coefficient of 370~mJ/K$^2$~$\cdot$~mol indicating the heavy fermion nature of CeMg$_3$.   An estimation of Kondo temperature $T_{\rm K}$ was made and found that it is of similar magnitude as that of $T_{\rm N}$.  The reduced value of the magnetization below the ordering temperature, together with the reduced entropy at the magnetic ordering temperature and the enhanced low temperature heat capacity indicates that Kondo effect plays a significant role in this compound. The electrical resistivity measurement suggests that CeMg$_3$ is a  Kondo lattice compound.   We have performed the crystalline electric field (CEF) analysis on the magnetic susceptibility and the heat capacity data and found that the ground state is a $\Gamma_7$ doublet with an overall splitting of 191~K.   

\end{abstract}



\maketitle
\section {Introduction}
The magnetism due to the localized $4f$ electrons in the rare-earth based intermetallic compounds is always interesting due to the wide range of physical properties exhibited by these compounds.  The binary  rare-earth intermetallic systems have been studied over several decades.  In  particular, the Ce based binary compounds are the most extensively studied as they exhibit many novel magnetic behavior owing to the close proximity of the $4f$  level to the Fermi level.  Like for example, CeAl$_3$ and CeCu$_6$ are heavy fermion Kondo lattice compounds without any long-range magnetic order~\cite{Andres, Onuki}, the cubic CeAl$_2$ and CeB$_6$ are Kondo lattice compounds that order magnetically at low temperature~\cite{Barbara, Winzer}.  Pressure induced superconductivity is observed in the binary compound CeIn$_3$~\cite{Grosche}.  In view of these interesting properties of binary compounds, we wanted to investigate the magnetic properties of RMg$_3$ (R = La and Ce) compounds which crystallize in the cubic BiF$_3$- type structure.   In one of the early reports on RMg$_3$ system, Buschow has studied the magnetic properties of CeMg$_3$ and NdMg$_3$ and reported no magnetic ordering, down to 4.2~K,  in these two compounds~\cite{Buschow}.  However, Galera \textit{et al} and Pierre \textit{et al} have found that  CeMg$_3$ orders antiferromagnetically at 3.4~K~\cite{Pierre, Galera}, while NdMg$_3$ orders magnetically at  6~K~\cite{Galera}.  From the previous neutron scattering experiments on polycrystalline sample~\cite{Galera2, Tapan}, it has been found that NdMg$_3$ undergoes an antiferromagnetic ordering at $T_{\rm N} = 6$~K with the propagation vector {\bf k}~=~(0.5, 0.5, 0.5) and the magnetic structure of NdMg$_3$ possesses ferromagnetic layers of (111) which are stacked in opposite direction leading to an antiferromagnetic alignment along the [111] axis.    Recently, Tanida \textit{et al} have performed a  detailed magnetic  measurements on PrMg$_3$ single crystal and found that PrMg$_3$ exhibits a  nonmagnetic $\Gamma_3$ doublet ground state~\cite{Tanida}.   In this paper, we report on the detailed magnetic properties of single crystalline LaMg$_3$ and CeMg$_3$ by studying the magnetization, electrical resistivity and heat capacity measurements. A crystal electric field (CEF) analysis was performed on the susceptibility and heat capacity data and the energy levels thus obtained corroborates the previous neutron diffraction measurements performed on polycrystalline samples~\cite{Pierre}.

\section{Experiment}

From the binary phase diagram of Ce-Mg and La-Mg by Nayeb-hasehmi and Clark~\cite{Nayeb} it was found that both CeMg$_3$ and LaMg$_3$ melt congruently at temperature close to 800~$^\circ$C. Hence these compounds can be grown directly from the melt.  Owing to the high vapor pressure of Mg,  the single crystals were grown by Bridgman method.  The stoichiometric amounts of the rare-earth metal and magnesium were taken in a point bottomed alumina crucible and sealed in a molybdenum tube.  This molybdenum tube was then subsequently sealed in a quartz ampoule and loaded into a box type furnace.  The furnace was then raised to 850~$^\circ$C, well above the melting point of these compounds and held at this temperature for 24~hours.  Then the samples were slowly cooled down to 780~$^\circ$C over a period of 3~days.  Bulk shiny single crystals were obtained by gently taping on the alumina crucible and they were found to be stable in air.  The crystals were then subjected to powder x-ray diffraction to check the phase purity. The orientation of the crystal was done by back reflection Laue  method.   The dc magnetic susceptibility and the magnetization measurements were performed in the temperature range 1.8-300~K using a superconducting quantum interference device (SQUID) and vibrating sample magnetometer (VSM).   The electrical resistivity was measured down to 1.9~K in a home made set up.  The heat capacity was measured using a Quantum Design physical property measurement system (PPMS).

\section{Results}
\subsection{X-ray diffraction}

\begin{figure}[h]
\includegraphics[width=0.4\textwidth]{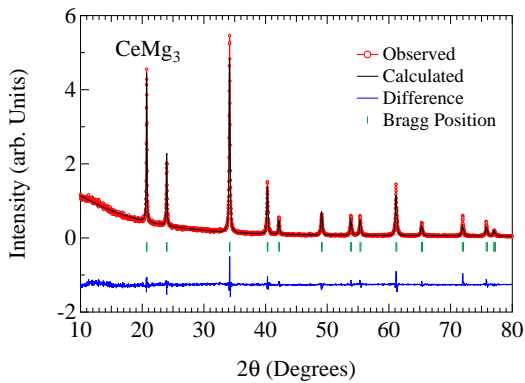}
\caption{\label{fig1}(Color online) A representative powder x-ray diffraction pattern of  CeMg$_3$. }
\end{figure}

Small pieces of the single crystals were crushed into fine powder and subjected to powder x-ray diffraction, using a PANalytical x-ray diffractometer with monochromatic Cu-K$_{\rm \alpha}$ radiation, to check the phase purity  and to estimate the lattice constant values.  No traces of any  secondary phases were seen in the x-ray diffractogram indicating the samples are phase pure. A Rietveld analysis was performed and the representative powder x-ray diffraction pattern of CeMg$_3$ is shown in Fig.~\ref{fig1}.   A reasonably good fit of the experimental pattern confirms the space group $Fm\bar{3}m$ (\#225) and the 
estimated lattice constant was 7.468~$\rm {\AA}$  for LaMg$_3$ and 7.422~$\rm {\AA}$  for CeMg$_3$.  It is worth mentioning here that the nearest Ce-Ce distance is 5.248~\AA~.  Hence, CeMg$_3$ is a good system to study the competing  Ruderman-Kittel-Kasuya-Yosida (RKKY) interaction and Kondo screening. The stoichiometry of the samples was further  confirmed by the energy dispersive analysis by x-ray (EDAX) where the composition of the single crystal was analysed at various regions of the single crystal.    It was found that the sample is highly homogeneous maintaining the stoichiometry through out the sample.  The crystals were then oriented along the principal crystallographic direction namely [100], by means of Laue back reflection.  The  four fold symmetry of the Laue pattern confirmed the cubic structure of this compound.  The crystals were then cut along [100] by means of a spark erosion cutting machine for the magnetic property measurements.

\subsection{Magnetization}

\begin{figure}[h]
\includegraphics[width=0.4\textwidth]{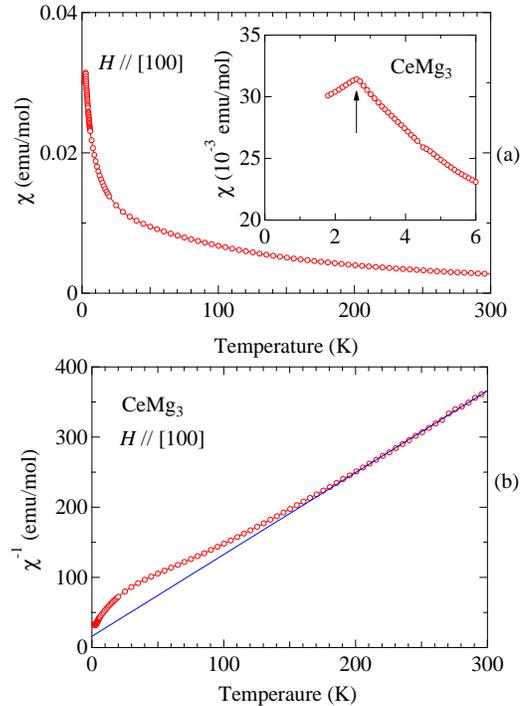}
\caption{\label{fig2}(Color online) (a) Temperature dependence of magnetic susceptibility of CeMg$_3$ for field parallel to [100].  The inset shows the low temperature part of the magnetic susceptibility in an expanded scale.  (b) Inverse magnetic susceptibility of CeMg$_3$, the solid line  is a fit to Curie-Weiss law.  }
\end{figure}

The temperature dependence of magnetic susceptibility of CeMg$_3$ from 1.8  to 300~K in an applied magnetic field of  1~kOe for $H$ parallel to [100] direction  is shown in Fig.~\ref{fig2}(a).  The inset
shows the low temperature part of the magnetic susceptibility where the susceptibility exhibits a cusp at 2.6~K and drops with the decrease in the temperature.  This cusp in the magnetic susceptibility at 2.6~K is the N\'{e}el temperature of CeMg$_3$ where the Ce moments order antiferromagnetically.  The magnetic ordering temperature at 2.6~K here is less than what it has been reported on polycrystalline sample earlier~\cite{Galera}. The heat capacity measurement to be discussed later confirms the bulk magnetic ordering at this temperature.  Figure~\ref{fig2}(b) shows the inverse magnetic susceptibility. At high temperature the inverse magnetic susceptibility is linear and follows Curie-Weiss law for temperature above 200~K.  The solid line in Fig.~\ref{fig2}(b) is a fit to the Curie-Weiss law.  From the fitting, effective magnetic moment $\mu_{\rm eff}$ and  the paramagnetic Curie temperature $\theta_{\rm p}$ were found to be 2.61~$\mu_{\rm B}$/Ce and -12~K, respectively.  The estimated effective moment is close to the free ion value of Ce, 2.54~$\mu_{\rm B}$ in its trivalent state.  The negative sign of the $\theta_{\rm p}$ indicates the antiferromagnetic nature of the magnetic ordering.  The inverse susceptibility deviates from the linearity for temperature less than 200~K which is mainly attributed to the crystal electric field (CEF) effect.

\begin{figure}[h]
\includegraphics[width=0.4\textwidth]{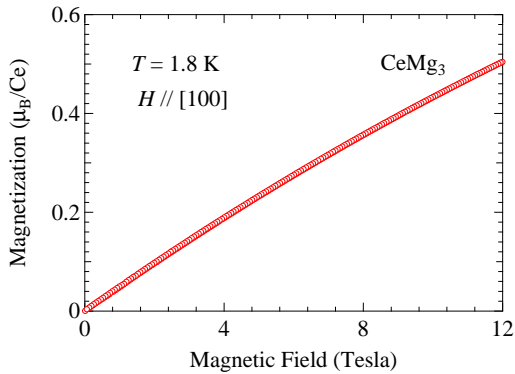}
\caption{\label{fig3}(Color online) Isothermal magnetization of CeMg$_3$ for $H~\parallel$~[100] at T = 1.8~K measured in a vibrating sample magnetometer.}
\end{figure}

The field dependence of magnetization $M(H)$ at a constant temperature $T$~=~1.8~K  is shown in Fig.~\ref{fig3}.  The magnetization is almost linear up to a field of 12~T, thus confirming the
 antiferromagnetic ordering of the Ce moments.   The magnetization does not show any signature of saturation up to a field of 12~T and attains a value of only 0.5~$\mu_{\rm B}$/Ce, thus corroborating the previous neutron diffraction experimental results where an ordered moment of 0.59~$\mu_{\rm B}$ is observed~\cite{Galera}.  The saturation moment of a free Ce$^{3+}$ ion $g_JJ\mu_{\rm B}$ is 2.14~$\mu_{\rm B}$ with $g_J$~=6/7 and $J$~=~5/2.  The observed magnetization is much reduced compared to the free Ce$^{3+}$ ion and this reduction in the magnetization is attributed to the Kondo effect and CEF effect.

\subsection{Electrical Resistivity}
\begin{figure}[h]
\includegraphics[width=0.4\textwidth]{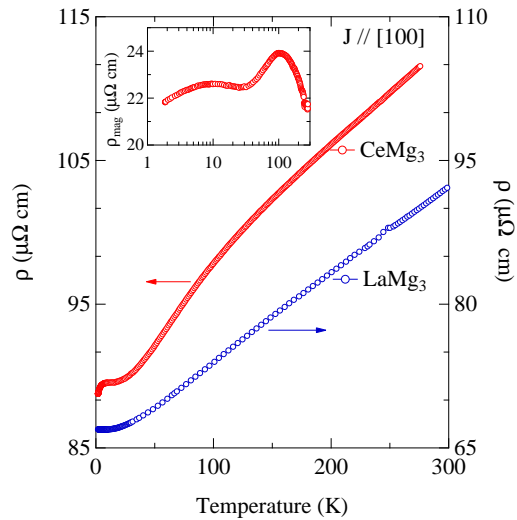}
\caption{\label{fig4}(Color online) (a) Temperature dependence of electrical resistivity $\rho (T)$ of CeMg$_3$ and LaMg$_3$ in the temperature range 1.9 to 300~K. Inset shows the magnetic part of the electrical resistivity $\rho_{\rm mag} (T)$.}
\end{figure}

The electrical resistivity of CeMg$_3$ and LaMg$_3$ in the temperature range from 1.9 to 300~K, for current parallel to [100] direction is shown in the main panel of Fig.~\ref{fig4}.  
The antiferromagnetic ordering at 2.6~K is not clearly discernible here as the resistivity was measured down only to 1.9~K.   The $\rho (T)$ of LaMg$_3$ decreases with the decrease in temperature, typical of a metallic compound without any anomaly at low temperature while the electrical resistivity for  CeMg$_3$ initially decreases with decreasing temperature and exhibits a broad curvature at around 150~K followed by a minimum at 30~K.   Below 30~K, the resistivity shows a weak increase and then decreases.  The broad hump at 150~K may be attributed to the combined influence of the crystal field and Kondo effects.   The inset shows the magnetic part of the resistivity  $\rho_{\rm mag}$, obtained by subtracting $\rho_{\rm LaMg_3}$ from  $\rho_{\rm CeMg_3}$, in semi-logarithmic scale.  $\rho_{\rm mag}$ increases with the decrease in temperature below 300~K and shows two broad peaks centered around 6~K and 100~K.  This type of double peaked structure in $\rho_{\rm mag} (T)$ is often observed in Kondo lattice compounds, exhibiting a magnetically ordered ground states.  Hence, CeMg$_3$ can be termed as a Kondo lattice compound. The $\rho_{\rm mag} (T)$ varies nearly as --ln(T) at low and high temperatures.  According to Cornut and Coqblin~\cite{Cornut}, this type of behaviour is expected for Kondo type interaction in the presence of strong crystal field splitting with the Kondo temperature $T_{\rm K}$ much less than the over all crystal field splitting $\Delta_{\rm CEF}$.  The low temperature peak in  $\rho_{\rm mag} (T)$ may be attributed to the Kondo scattering of the conduction electrons with an energy scale of the order of Kondo temperature $T_{\rm K}$.  The estimation of $T_{\rm K}$ is given in the discussion part.

\begin{figure}[h]
\includegraphics[width=0.4\textwidth]{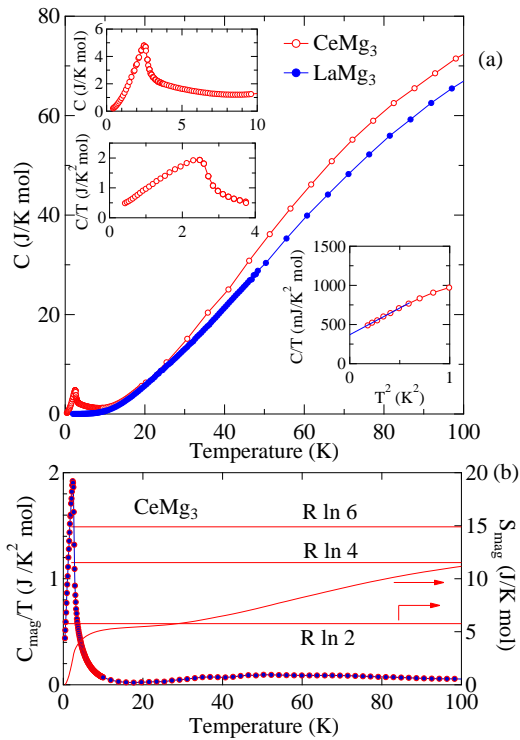}
\caption{\label{fig5}(Color online) (a) Temperature dependence of the specific heat capacity in CeMg$_3$ and LaMg$_3$.  The top insets show the low temperature plots of C vs. T and C/T vs. T of CeMg$_3$.  The bottom inset shows the C/T vs T$^2$ plot.  (b)  C$_{\rm mag}$/T vs. T of CeMg$_3$.  The calculated entropy is plotted in the right axis.}
\end{figure}
\subsection{Heat Capacity}

 The temperature dependence of the specific heat capacity of single crystalline CeMg$_3$ and  LaMg$_3$ are shown in the main panel of Fig.~\ref{fig5}(a). The heat capacity of LaMg$_3$ does not show any anomaly, and its temperature dependence is typical for a non-magnetic reference compound.  The low temperature part of the heat capacity of CeMg$_3$ is shown in the top inset of Fig.~\ref{fig5}(a), where a clear jump is seen at  $T_{\rm N}$ = 2.6~K, confirming the bulk magnetic ordering in this compound.  No anomaly is seen at 3.5 or 4~K as claimed by Galera et al~\cite{Galera} in the earlier studies on polycrystalline samples.   Thus the antiferromagnetic ordering in CeMg$_3$ is confirmed to be at 2.6~K.  The low temperature part of the  $C/T$ vs. $T^{2}$ plot is shown in Fig.~\ref{fig5}(a) and the magnitude of the Sommerfeld coefficient $\gamma$ is obtained by fitting the expression $C/T = \gamma + \beta T^2$.  The  $\gamma$ value thus obtained was estimated to be 370~mJ/K$^2\cdot$mol, and the $\beta$ value is estimated to be 608~mJ/K$^{4} \cdot $ mol.    The high temperature $\gamma_{\rm HT}$ estimated in the paramagnetic state (12~K~\textless~T~\textless~17~K) results in a value of 41~mJ/K$^2 \cdot$ mol.  This implies that the enhanced low temperature $\gamma$ value is due to the strong Kondo interaction.   The large value of  $\gamma$ in the low temperature region indicates that  CeMg$_3$ is a heavy fermion compound, while the obtained $\beta$ value implies the expected contribution from the magnons, in the magnetic structure.   The large value of the $\gamma$ also signals the enhanced density of quasi-particle state at the Fermi level. The magnetic part of heat capacity $C_{\rm mag}$ was estimated by the usual method of subtracting the heat capacity of LaMg$_3$ from  CeMg$_3$.  $C_{\rm mag}$ shows a broad peak centered around 80~K indicating the Schottky anomaly.  The energy level scheme is given in the discussion part.  The $C_{\rm mag}$/T versus temperature plot and the calculated entropy is shown in Fig.~\ref{fig5}(b).  At $T_{\rm N}$ (= 2.6~K) the entropy amounts to only $0.5 R~\rm{ln~2}$ with respect to the value of $R~{\rm ln~2}$ anticipated for a doublet ground state.  The reduced value of  $R~{\rm ln~2}$  further confirms the Kondo effect in CeMg$_3$.

\begin{figure}[h]
\includegraphics[width=0.4\textwidth]{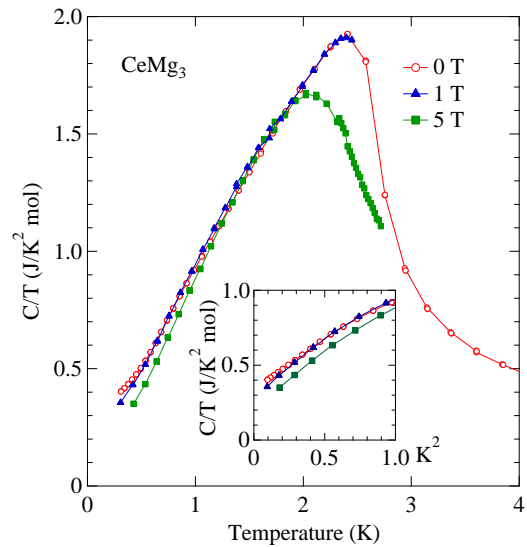}
\caption{\label{fig6}(Color online) (a)  C/T versus T plot of CeMg$_3$ in applied magnetic fields, in the temperature range 0.5 to 4~K.  The inset shows the low temperature part of C/T versus T$^2$ plot.  }
\end{figure}
 
 Figure~\ref{fig6} shows the $C/T$ versus $T$ plot of CeMg$_3$ measured in applied fields of 0,  1 and 5~T.  There is no appreciable change in the ordering temperature in an applied field of 1~T while in a 5~T field, the N\'{e}el temperature shifts to lower temperature and the jump in the heat capacity also decreases as it is usually observed in antiferromagnetically ordering compounds. The low temperature part of $C/T$ versus $T^2$ plot is shown in the inset of Fig.~\ref{fig6}. An estimation of the $\gamma$ value by linear extrapolation of the $C/T$ versus $T^2$ plot of the heat capacity data measured in 1~T and 5~T field results in 300 and 200~mJ/K$^2\cdot$mol, respectively.  The reduction in the $\gamma$ value implies that the application of magnetic field tends to break the Kondo coupling between the localized $4f$ electron and the conduction electrons.  
  
\section{Discussion}

\begin{figure}[h]
\includegraphics[width=0.4\textwidth]{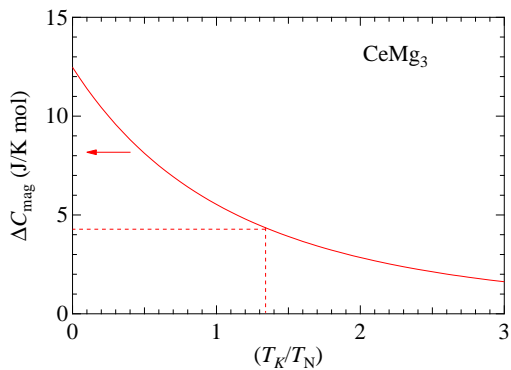}
\caption{\label{fig7} (Color online) Estimation of the Kondo temperature from the jump in the magnetic part of the heat capacity.  The solid line is the plot of Eq.~\ref{eqn1}.  See text for details. }
\end{figure}

The previous studies on polycrystalline samples of CeMg$_3$ have claimed that this compound orders antiferromagnetically at 4~K~\cite{Galera}.  However, it is ubiquitous from our magnetic and thermal measurements on  clean phase pure single crystals that CeMg$_3$ orders antiferromagnetically at 2.6~K. The large value of the Sommerfeld coefficient $\gamma$ at low temperature indicates enhanced density of states at the Fermi level which is due to the result of Kondo effect.  The electrical resistivity measurement clearly indicates that CeMg$_3$ is a Kondo lattice compound.   The reduced value of the magnetic moment of  0.5~$\mu_{\rm B}$/Ce at $T$~=~1.8~K in a field of 12~T  and a very small heat capacity jump at the magnetic ordering temperature further confirms the Kondo effect  in CeMg$_3$.  We made an estimate of the Kondo Temperature $T_{\rm K}$ by the method described by Bredl \textit{et al.}~\cite{Bredl}, where in the mean field approach, the jump in the heat capacity  $\Delta C_{\rm mag}$ of a Kondo system is related to the Kondo temperature.  Besnus \textit{et al}~\cite{Besnus} have found that the estimation of $T_{\rm K}$ by this model agrees well with the experimental data  on various Ce and Yb compounds.  From Eqs.~(12), (13) and (15) of Ref.~\onlinecite{Bredl}, Blanco~\textit{et al}. have given the expression for $\Delta C_{\rm mag}$  at the magnetic ordering temperature  as,
\begin{equation}
\label{eqn1}
\Delta C_{\rm mag} = \frac{6N_{\rm A} k_{\rm B}}{\psi^{'''}\left(\frac{1}{2} + x \right)}\left[\psi^{'}\left(\frac{1}{2} + x \right) + x \psi^{''} \left(\frac{1}{2} + x \right)\right]^2,
\end{equation}
where $x~=~(T_{\rm K}/T_{\rm N})/2\pi$ and $\psi^{'}, \psi{''}$ and $\psi^{'''}$ are the first, second and third derivative of the polygamma function and the other terms have the usual meaning.  A plot of this function is shown in Fig~\ref{fig7}.  The jump in the magnetic part of the heat capacity was estimated to be 4.35~J/K~$\cdot$~mol, which results in  a $T_{\rm K}$  value of 3.5~K.  Desgranges and Schotte~\cite{Desgranges} have 
theoretically explained that the entropy of a Kondo system at the characteristic temperature $T_{\rm K}$ amounts to 0.68~$R$~ln~2 (= 3.92~J/K~$\cdot$~mol).  According to this expression, for CeMg$_3$,  0.68~$R$~ln~2 is reached at 3.8~K which indicates that this Kondo temperature is in close agreement with the one estimated from  Eq.~\ref{eqn1}.  Furthermore, the Kondo temperature can also be estimated from the paramagnetic Curie-Weiss temperature as mentioned by  Gruner~\textit{et al}~\cite{Gruner}, $|\theta_{\rm p}|/4~ (12/4 =  3$~K).  The estimation of Kondo temperature $T_{\rm K}$ by various methods gives a value of 3-4~K, which is close to the magnetic ordering temperature of CeMg$_3$.  However, a detailed neutron diffraction experiment has to be performed to substantiate our estimation of T$_{\rm K}$.

\begin{figure}[!]
\includegraphics[width=0.4\textwidth]{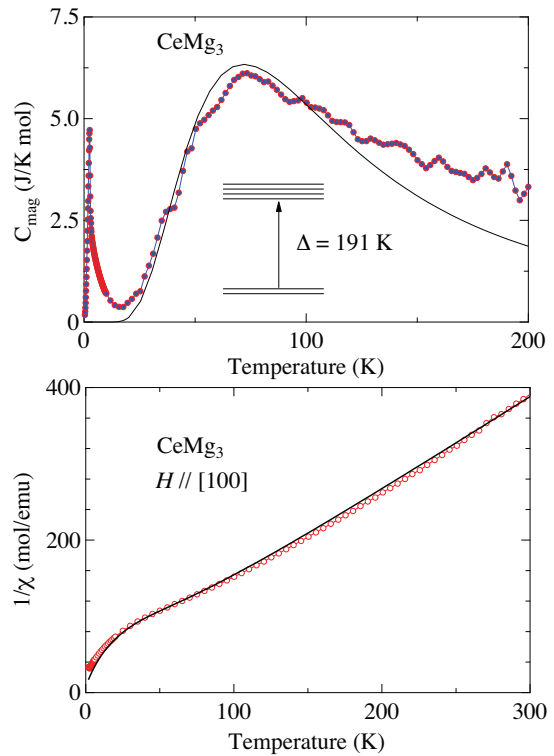}
\caption{\label{fig8} (Color online) (a) Magnetic part of the specific heat capacity of CeMg$_3$.  The solid line is calcualted Schottky heat capacity, which gives an energy separation of 191~K between the ground state doublet and the excited quartet.  (b) Inverse magnetic susceptibility of CeMg$_3$.  The solid line is based on the CEF calculation.  }
\end{figure}

The magnetic part of the heat capacity $C_{\rm mag}$ of CeMg$_3$  shows a broad peak at high temperature as shown in Fig.~\ref{fig8}(a).   This feature is attributed to the Schottky excitations between the CEF levels of the Ce$^{ 3+}$ ions.  We have performed an analysis of the heat capacity  and the magnetic susceptibility data on the basis of CEF model.   The Schottky contribution to heat capacity is given by the following expression,
\begin{equation}
\label{eqn2}
C_{Sch}\left(T\right)=R\left[\frac{\displaystyle\sum\limits_{i}g_{i}e^{-E_{i}/T}\sum_{i}g_{i}E_{i}^{2}e^{-E_{i}/T}-\left[\sum_{i}g_{i}E_{i}e^{-E_{i}/T}\right]^{2}}{T^{2}\left[\displaystyle\sum\limits_{i}g_{i}e^{-E_{i}/T}\right]^{2}}\right],
\end{equation}
where $R$ is the gas constant, E$_{\mathrm{i}}$ is the CEF energy level in units of temperature and g$_{\mathrm{i}}$ the corresponding degeneracy.  The Ce atom in CeMg$_3$ possesses a cubic site symmetry and hence the ground state should be a doublet or quartet.  From the entropy calculations discussed above, it is obvious that the ground state of CeMg$_3$ should be a doublet ground state.  Considering a degeneracy of doublet ground state and a quartet excited state, we found that the CEF levels are separated by an energy of 191~K apart.  The solid line in Fig.~\ref{fig8}(a) indicates the calculated Schottky heat capacity which is matching well with the experimental data.  

The crystalline electric field analysis was also performed on the magnetic susceptibility data.  For a cubic point symmetry, the CEF Hamiltonian is given by,
\begin{equation}
\label{eqn3}
\mathcal{H}_{\rm CEF} = B_{4}^{0}\left(O_{4}^{0}+5O_{4}^{4}\right)+B_{6}^{0}\left(O_{6}^{0}-21O_{6}^{4}\right)
\end{equation},
where $B_{l}^{m}$ and $O_{l}^{m}$ are the crystal field parameters and the Steven's operators respectively~\cite{Stevens, Hutchings}.  For Ce atom, the sixth order terms $O_{6}^{0}$ and $O_{6}^{4}$ vanish and hence the CEF Hamiltonian reduces to 

\begin{equation}
\label{eqn4}
\mathcal{H}_{\rm CEF} = B_{4}^{0}\left(O_{4}^{0}+5O_{4}^{4}\right)
\end{equation}.

 The CEF susceptibility is given by the expression,
\begin{widetext}
\begin{equation}
\label{eqn5}
\chi_{{\rm CEF}i} = N(g_{J}\mu_{\rm B})^2 \frac{1}{Z}
\left(\sum_{m \neq n} \mid \langle m \mid J_{i} \mid n \rangle
\mid^{2} \frac{1-e^{-\beta \Delta_{m,n}}}{\Delta_{m,n}}e^{-\beta
E_{n}} +  \sum_{n} \mid \langle n \mid J_{i} \mid n \rangle
\mid^{2} \beta e^{-\beta E_{n}} \right),
\end{equation}
\end{widetext}
where $g_{J}$ is the Land\'{e} $g$\,-\,factor, $E_{n}$ and $\mid\!n \rangle$ are the $n$th eigenvalue and eigenfunction, respectively.  $J_{i}$ ($i$\,=\,$x$, $y$ and $z$) is a component of the angular momentum,  and
$\Delta_{m,n}\,=\,E_{n}\, - \,E_{m}$, $Z\,=\,\sum_{n}e^{-\beta
E_{n}}$ and $\beta\,=\,1/k_{\rm B}T$.  The magnetic susceptibility
including the molecular field contribution $\lambda_{i}$ is given
by
\begin{equation}
\label{eqn6}
\chi^{-1}_{i} = \chi_{{\rm CEF}i}^{-1} - \lambda_{i}.
\end{equation}

The CEF parameter $B_{4}^{0}$ was estimated by using the Eqns.~\ref{eqn4}, ~\ref{eqn5} and \ref{eqn6}.  The solid line Fig.~\ref{fig8}(b) is the calculated CEF curve  and the value of $B_{4}^{0}$ thus obtained is 0.53~K and the molecular field contribution $\lambda$ was $\--8$~mol/emu.  The sign of the $B_{4}^{0}$ parameter is positive, which clearly indicates that the ground state is a $\Gamma_7$ doublet.  From the eigen values of the crystal field Hamiltonian the crystal field splitting energy was found to be 191~K.   The crystal field level scheme thus obtained from the magnetic susceptibility and the heat capacity data are in accordance with the previous neutron diffraction results on polycrystalline samples~\cite{Pierre, Galera2}. The magnitude of the ordered moment of a $\Gamma_7$ doublet ground state based on the CEF calculation ($g_J J_x = 6/7~\times~0.833)$ should be 0.714~$\mu_{\rm B}$/Ce.  The observed magnetization value at 1.8~K is only 0.56~$\mu_{\rm B}$/Ce, this confirms that the moment reduction in this compound is due to the combined effect of Kondo and crystal field effects.  

The overall magnetic behaviour of CeMg$_3$ resembles to that of the popular antiferromagnetically ordered heavy fermion systems like CeIn$_3$~\cite{Knebel}, CePd$_2$Si$_2$~\cite{Grosche} and CeCu$_2$Ge$_2$~\cite{Jaccard}.  These systems have their N\'{e}el and Kondo temperatures nearly equal and exhibit pressure induced superconductivity.  The ordered moment in these compounds are also considerably reduced, and these compounds also possess enhanced Sommerfeld coefficient $\gamma$, as it has been observed in the present CeMg$_3$ system.  All these implies that CeMg$_3$ is another  compound where the high pressure studies will yield some interesting results.  

\section{Conclusion}

Single crystals of LaMg$_3$ and CeMg$_3$ were  grown by  Bridgman method in sealed molybdenum tubes.  X-ray and EDAX analysis of the sample confirmed the stoichiometry of the single crystals.  The magnetic susceptibility and the heat capacity clearly indicated the antiferromagnetic ordering at $T_{\rm N}$~=~2.6~K.   The heavy fermion nature of this compound is confirmed by the large  Sommerfield coefficient $\gamma$~=~370 mJ/K$^2$~$\cdot$~mol.  The reduced value of the  magnetization and the reduced magnetic entropy at the magnetic ordering temperature  and the --ln(T) behaviour in the electrical resistivity confirmed the significant contribution of the Kondo effect in CeMg$_3$.  The estimated Kondo temperature $T_{\rm K}$ was found to be  3-4~K which is close to the magnetic ordering temperature, suggesting a delicate competition between the Kondo effect and the indirect exchange interaction.  Since the $T_{\rm N}$ and $T_{\rm K}$ are  very close in CeMg$_3$, it will be interesting to study the effect of pressure on the electrical resistivity of CeMg$_3$, which is planned for the future. The CEF analysis of the heat capacity and the magnetic susceptibility data indicated that the ground state is a $\Gamma_7$ doublet while the excited state is a $\Gamma_8$ quartet with an energy splitting of 191~K.

\section{Acknowledgement}

The discussions with Prof. S. K. Dhar is gratefully acknowledged.

\end{document}